\documentclass[preprintnumbers,amsmath,amssymb,floatfix,superscriptaddress,twocolumn,showpacs,pra]{revtex4}

\usepackage{epsfig}
\usepackage{graphicx} 
\usepackage{dcolumn}  
\usepackage{bm}       
\usepackage{color}
\usepackage{epsfig}
\usepackage{subfigure}
\usepackage{epstopdf}

\newcommand{\be}{\begin{equation}}
\newcommand{\ee}{\end{equation}}

\newcommand{\paper}{paper}

\newcommand{\ket}[1]{| #1 \rangle}

\newcommand{\matx}[1]{| #1 \rangle \langle #1 |}

\begin{document}
\title{Violation of multipartite Bell inequalities with classical subsystems via operationally local transformations}

\author{Mark S. Williamson} \email{cqtmsw@nus.edu.sg}
\affiliation{Centre for Quantum Technologies, National University of Singapore, 3 Science Drive 2, Singapore, 117543.}
\author{Libby Heaney} \email{l.heaney1@physics.ox.ac.uk}
\affiliation{Clarendon Laboratory, University of Oxford, Parks Road, Oxford, OX1 3PU, United Kingdom.}
\author{Wonmin Son}\email{cqtws@nus.edu.sg}
\affiliation{Centre for Quantum Technologies, National University of Singapore, 3 Science Drive 2, Singapore, 117543.}

\begin{abstract}
Recently, it was demonstrated by Son et al., Phys. Rev. Lett. \textbf{102}, 110404 (2009), that a separable bipartite continuous variable quantum system can violate the Clauser-Horne-Shimony-Holt (CHSH) inequality via operationally local transformations. Operationally local transformations are parameterized only by local variables, however in order to allow violation of the CHSH inequality a maximally entangled ancilla was necessary. The use of the entangled ancilla in this scheme caused the state under test to become dependent on the measurement choice one uses to calculate the CHSH inequality thus violating one of the assumptions used in deriving a Bell inequality, namely the `free will' or `statistical independence' assumption. The novelty in this scheme however is that the measurement settings can be external free parameters. In this paper we generalize these operationally local transformations for multipartite Bell inequalities (with dichotomic observables) and provide necessary and sufficient conditions for violation within this scheme. Namely, a violation of a multipartite Bell inequality in this setting is contingent on whether an ancillary system admits any realistic, local hidden variable model i.e. whether the ancilla violates the given Bell inequality. These results indicate that violation of a Bell inequality performed on a system does not necessarily imply that the system is nonlocal. In fact the system under test may be completely classical. However, nonlocality must have resided somewhere, this may have been in the environment, the physical variables used to manipulate the system or the detectors themselves provided the measurement settings are external free variables.
\end{abstract}

\maketitle

\section{Introduction}

Bell \cite{ref:Bell64} and others \cite{ref:CHSH69} formulated inequalities allowing experimental refutation of local, realistic models. Since then, experiments on two
\cite{ref:Aspect82a,ref:Tittel98,ref:Weihs98,ref:Rowe01,ref:Matsukevich08},
or more \cite{ref:Pan00,ref:Zhao03} entangled parties have consistently demonstrated that quantum mechanics is in disagreement with local realist predictions under reasonable assumptions.  In order to perform such Bell tests, a background reference frame is usually required to ensure phase locked conditions between both parties \cite{ref:Bartlett07}. To resolutely conclude nonlocality of a system, it is vital that this background reference admits a local hidden variable description, so that it is not the actual source of violation of the Bell inequality (see \cite{ref:Paterek10} for a particular example).

Until recently, tests of nonlocality have focused primarily on entangled systems, such as maximally entangled pairs of photons, since the strong correlations of entangled states are a necessary (but not sufficient) condition for the refutation of a local realistic model description. However, a recent paper by one of us \cite{ref:Son09} indicated that two space-like separated continuous variable systems were capable of violating the
Clauser-Horne-Shimony-Holt (CHSH) inequality \cite{ref:CHSH69} even when
the state was separable, i.e. describable by a local hidden variable theory.  However, in order to see this violation, a maximally entangled ancilla was \emph{a priori}
distributed between the parties and operational local transformations were applied between the two.   This ancilla could be, for instance, the background reference mentioned above.  

In this \paper\,, we will generalize the idea of such operationally local transformations (OLTs) to the
multipartite domain and provide necessary and sufficient conditions for violation of the family of Bell
inequalities with any number of dichotomic observable settings. A complete family of Bell inequalities with the added constraint of two observable settings are known; these are the Werner-Wolf-\.{Z}ukowski-Brukner (WWZB) inequalities \cite{ref:Werner&Wolf01,ref:Zukowski&Brukner02}. Our results apply to this and a wider family using the particular OLTs first introduced in \cite{ref:Son09} (hereafter known as SKKVB following the authors initials). Our findings indicate that the violation of a Bell inequality performed on a system does not necessarily imply that the system in question is nonlocal. In fact the system under test may be completely classical. The price to pay however, is that the state under test is no longer independent of the measurement settings violating the assumption of `free will' or `statistical independence' made in deriving a Bell inequality. In the scheme of SKKVB an ancilla mediates this measurement setting dependence and we show that it is necessary for this ancilla to violate a Bell inequality in order to observe a violation of a Bell inequality of the system under test. This ancilla could be regarded as entanglement in the background reference frame, between the physical variables used to manipulate the system or the detectors themselves.

The structure of the \paper\ is as follows: First we review the main findings of SKKVB and offer some clarifying observations on this scheme in section~\ref{sec:SKKVB}. Following this we give our main result in section~\ref{sec:mainresult}. We then give examples of specific states violating multipartite Bell inequalities in this scheme in section~\ref{sec:examples}. We also analyze the type of entanglement in SKKVBs scheme in section~\ref{sec:clusterstate} before concluding.

\section{Operationally local transformations}\label{sec:SKKVB}

SKKVB demonstrate that one can violate the CHSH inequality using a classically correlated state with positive Wigner function when particular OLTs are applied. The resource for the violation is a maximally entangled state shared between the parties which enables a larger class of OLTs to be performed.

We first introduce the CHSH inequality (a special case of WWZB) constructed from the expectation values
$\langle A^{(\alpha)} B^{(\beta)} \rangle=\text{tr}[A^{(\alpha)} B^{(\beta)} \rho_{ab}]$ of the observables $A^{(\alpha)}$ and $B^{(\beta)}$ whose single measurement outcomes made on each of the two quantum objects
labelled $a$ and $b$ are $\pm 1$. The CHSH inequality reads
\begin{equation}
|\langle A^{(1)} B^{(1)} \rangle + \langle A^{(1)} B^{(2)} \rangle + \langle A^{(2)} B^{(1)} \rangle - \langle A^{(2)} B^{(2)} \rangle | \leq 2.
\end{equation}
To determine the non-existence of a local hidden variable model describing $\rho_{ab}$, one looks for measurement settings, two for each party, $A^{(1)}$, $A^{(2)}$ and $B^{(1)}$, $B^{(2)}$, that violate
the upper bound of two.

 There exist two equivalent ways to perform the different measurements, $A^{(\alpha)}$ and $B^{(\beta)}$: 
(i) Rotate the measuring
devices and leave the state, $\rho_{ab}$, fixed.  If the measurement devices are initially in the $z$ basis (described by the Pauli $z$ operator $\sigma^{(3)}$=$\text{diag}\{1,-1\}$), they transform as
 $A^{(\alpha)}=R^{(\alpha) \dag} \sigma^{(3)}_a R^{(\alpha)}$ and $B^{(\beta)}=S^{(\beta) \dag} \sigma^{(3)}_b S^{(\beta)}$
where $R^{(\alpha)}$ and $S^{(\beta)}$ are general rotations (elements of the group $SU(2)$) acting on parties $a$ and
$b$ respectively. (ii) The second equivalent picture is to rotate the state, $\tilde{\rho}_{ab}=R^{(\alpha)} S^{(\beta)}
\rho_{ab}(R^{(\alpha)} S^{(\beta)})^\dag$, itself, while leaving the measurements, $\sigma^{(3)}_a$ and $\sigma^{(3)}_b$,
fixed. In the following, we will use picture (ii).

SKKVBs protocol starts with the initial, classically correlated state of the system
$\rho_{ab}=1/2(\matx{00}+\matx{11})$ and the maximally entangled ancilla state
$\chi_{a'b'}=\matx{\Phi^+}$, where $\ket{\Phi^+}_{a'b'}=1/\sqrt{2}(\ket{00}+\ket{11})$. Qubits $a$ and $a'$ are
given to the first party and qubits $b$ and $b'$ are given to the second space-like
separated party. Since SKKVB consider continuous variable systems, $\rho_{ab}$ consists of a classically correlated mixture of coherent states, $\ket{\alpha}$ and
$\ket{-\alpha}$.
In the present work, however, we map $\ket{\alpha}\rightarrow \ket{0}$ and $\ket{-\alpha}\rightarrow\ket{1}$. This mapping is isomorphic, since SKKVB choose $\alpha$ to be large enough to ensure $\langle \alpha | -\alpha\rangle \approx 0$. In the same way, their
measurements are fixed so that $A\rightarrow \sigma^{(3)}$. We choose this equivalent representation of SKKVB as we find it simpler.

To obtain the necessary degrees of freedom to construct the CHSH inequality, the state,  $\rho_{ab}$, is rotated. These extra degrees of freedom are encoded in the unitary OLTs
$U_{aa'}(\theta^{(\alpha)}_a)$ and $U_{bb'}(\theta^{(\beta)}_b)$ parameterized with local angles $\theta^{(\alpha)}_a$
and $\theta^{(\beta)}_b$ and are applied to qubits $a$ and $a'$ and $b$ and $b'$ respectively. Their form (in the computational basis $\{\ket{00}, \ket{01}, \ket{10}, \ket{11}\}$) reads
\begin{equation}
\label{eq:unitary}
U_{aa'}(\theta^{(\alpha)}_a)=\begin{pmatrix}
                  c^{(\alpha)}_a & -s^{(\alpha)}_a & 0 & 0 \\
                   0 & 0 &s^{(\alpha)}_a & c^{(\alpha)}_a\\
                   0 & 0 & c^{(\alpha)}_a & -s^{(\alpha)}_a \\
                  s^{(\alpha)}_a & c^{(\alpha)}_a & 0 & 0 \\
                 \end{pmatrix},
\end{equation}
and likewise for  $U_{bb'}(\theta^{(\beta)}_b)$, where $c^{(\alpha)}_a\equiv\cos\left(\theta^{(\alpha)}_a/2\right)$ and $ s^{(\alpha)}_a\equiv\sin\left(\theta^{(\alpha)}_a/2\right)$.  Such unitary
operators are defined as operationally local since there is no exchange of the local variables between the parties.

The state resulting from application of these OLTs is
$\varrho_{aba'b'}(\theta^{(\alpha)}_a,\theta^{(\beta)}_b)=U_{aa'}U_{bb'} \rho_{ab}\chi_{a'b'}(U_{aa'} U_{bb'})^\dag$
giving the reduced state seen by $a$ and $b$ as $\rho_{ab}(\theta^{(\alpha)}_a,\theta^{(\beta)}_b)=\text{tr}_{a'b'}
[\varrho_{aba'b'} (\theta^\alpha_a,\theta^\beta_b)]$. One now performs a Bell test on this reduced state by
calculating the expectation values $\langle A^{(\alpha)} B^{(\beta)} \rangle=\text{tr}\left[\sigma^{(3)}_a \sigma^{(3)}_b \rho_{ab}(\theta^{(\alpha)}_a,\theta^{(\beta)}_b)\right]$.

For different values of the local angles $\theta^{(\alpha)}_a$ and $\theta^{(\beta)}_b$, two for each angle,
one can now construct the CHSH inequality. For the values $(\theta^{(1)}_a,\theta^{(2)}_a,\theta^{(1)}_b,\theta^{(2)}_b)=(0,\pi/2,\pi/4,-\pi/4)$
one finds the maximal violation of $2\sqrt{2}$ even though the state remains separable (and has a positive Wigner function). Namely the state is $\rho_{ab}(\theta^{(\alpha)}_a,\theta^{(\beta)}_b)=
1/2[(1+\cos(\theta^{(\alpha)}_a-\theta^{(\beta)}_b))\rho_{ab}+ (1-\cos(\theta^{(\alpha)}_a-\theta^{(\beta)}_b))\bar{\rho}_{ab}]$,
a mixture of the perfectly classically correlated, $\rho_{ab}$, and anti-correlated, $\bar{\rho}_{ab}=1/2(\matx{01}+\matx{10})$,
states.

\subsection{Discussion}

Note that $\rho_{ab}(\theta^{(\alpha)}_a,\theta^{(\beta)}_b)$ is now dependent on the local free parameters of the measurement choice, namely $\theta^{(\alpha)}_a$ and $\theta^{(\beta)}_b$. Because of this dependence, the assumption of `free will' or `statistical independence' assumed in deriving a Bell inequality no longer holds. In this scheme the dependence is mediated by the entanglement in the ancilla. That is, the state the Bell inequality is performed on is dependent on the measurement settings prior to the measurement being made \cite{ref:Bell77}. It should be noted that one may violate a Bell inequality with only classical correlations when giving up the assumption of statistical independence if one supplements the state with extra degrees of freedom encoded in an extra local `hidden' variable that one can measure. The novelty in SKKVBs scheme however is that the choices of the measurement settings can be external parameters chosen randomly. In the purely classical scheme the measurement choices are determined by the results of measuring the local, hidden variable.

We may also view this scheme in a way that does not violate the statistical independence assumption. One can view the system state $\rho_{ab}$ and the OLTs as part of a positive operator valued measurement (POVM). Projective measurements are a special case of these more general measurements. One way of performing a POVM is to take an ancilla and perform a global unitary operation on this ancilla and the system one wishes to measure. Following the global operation one performs a projective measurement on the ancilla. In the scheme described above, if one thinks of the state $\rho_{ab}$ as the ancilla, the state $\chi_{a'b'}$ as the state one wishes to measure and $U_{aa'}$ and $U_{bb'}$ as the global operations it becomes clear that the entangled state $\chi_{a'b'}$ is the system the Bell test is being performed on and of course this must be entangled to see a violation of the Bell inequality. Viewed in this way, the result of SKKVB can be clarified. However, SKKVB show that if one regards $\rho_{ab}$ as the system under test then it appears one can violate a Bell inequality with a separable state. This point is important when considering experimental tests of Bell inequalities.

Bell conjectured positivity of the Wigner function when performing phase-space measurements should not allow violation of a
Bell inequality. SKKVB showed that this statement should be taken with care particularly when deciding which is the system under test. OLTs enable the possibility of a violation since the system state becomes correlated to the measurement parameters via the maximally entangled ancilla.

The question we address in the following section is under which conditions is it
possible to observe a violation of a multipartite Bell inequality with dichotomic observables within SKKVBs scheme? For example, can one still violate the CHSH inequality if the ancilla
is not maximally entangled? We will give necessary and sufficient conditions for both initial system state $\rho_{ab}$
and ancilla $\chi_{a'b'}$ for a violation to be observed. The proof is extended beyond the bipartite scenario to multi-qubit schemes and we will show
necessary and sufficient conditions for this more general setting, that is, all multi-party Bell inequalities with an unlimited number of dichotomic observable settings \footnote{For an example of a Bell inequality with more than two dichotomic settings that identify states not admitting a local hidden variable model and not violating CHSH see \cite{ref:Collins&Gisin04}.}.

\section{Necessary and sufficient conditions for violation of Bell inequalities with dichotomic observables using OLTs}\label{sec:mainresult}

Consider now a generalization of SKKVBs scheme to a many party qubit Bell test. One starts with the system state $\rho_{a b \ldots z}$ (which can be separable)
consisting of qubits $a, b,\ldots, z$, each of which is distributed to space-like separated parties. We also distribute a second qubit $a', b', \ldots, z'$ of ancilla state $\chi_{a' b' \ldots z'}$ to each of these parties. Each party now performs
the unitaries $U_{a a'}(\theta_a^{(\alpha)}), U_{b b'}(\theta_b^{(\beta)}),\ldots, U_{zz'}(\theta_z^{(\omega)})$ between their ancilla
qubit and system qubit for some value of their chosen local parameter $\theta^{(\alpha)}$. These two qubit unitaries are
defined analogously to those given in eq.~(\ref{eq:unitary}). The resulting overall state is therefore $\varrho_{ab\ldots z a' b' \ldots z'}
(\theta^{(\alpha)}_a,\theta^{(\beta)}_b,\ldots,\theta^{(\omega)}_z)=U_{aa'}U_{bb'}\ldots U_{zz'}\rho_{a b \ldots z}\chi_{a' b' \ldots z'}
(U_{aa'} U_{bb'}\ldots U_{zz'})^\dag$ with reduced state $\rho_{a b \ldots z}(\theta^{(\alpha)}_a,\theta^{(\beta)}_b,\ldots,\theta^{(\omega)}_z)$
found by tracing over the ancilla.

One performs the Bell test in fixed measurement basis $\sigma^{(3)}_a \sigma^{(3)}_b \ldots \sigma^{(3)}_z$
on this reduced state yielding the expectation values
\begin{eqnarray}\label{eq:expectationvalN}
&&\langle A^{(\alpha)} B^{(\beta)} \ldots Z^{(\omega)} \rangle = \nonumber\\ &&\text{tr}\left[\sigma^{(3)}_a \sigma^{(3)}_b \ldots \sigma^{(3)}_z \rho_{a b \ldots z}
(\theta^{(\alpha)}_a,\theta^{(\beta)}_b,\ldots,\theta^{(\omega)}_z)\right].
\end{eqnarray}
Having introduced this multiparty generalization we can now announce the following:

\emph{Theorem:} The state $\rho_{a b \ldots z}(\theta^{(\alpha)}_a,\theta^{(\beta)}_b,\ldots,\theta^{(\omega)}_z)$ violates a Bell inequality with an unlimited number of dichotomic observable settings by some amount if and only if (iff)
the ancilla state $\chi_{a'b'\ldots z'}$ also violates the same inequality by the same amount \emph{provided} the initial state $\rho_{ab\ldots z}$
is an eigenstate of $\sigma^{(3)}_a \sigma^{(3)}_b \ldots \sigma^{(3)}_z$.

\emph{Proof:} The key ingredient in this proof is a convenient decomposition of the unitaries into the product of a
controlled-NOT (CNOT) transformation with a local rotation on the ancilla qubit. That is,
\begin{equation}
U_{aa'}(\theta_a^{(\alpha)})=C_{a'a}R_{a'}(\theta_a^{(\alpha)})
\end{equation}
where $C_{a'a}$ is the CNOT gate, the first subscript labeling the control and the second the target qubit. $R_{a'}(\theta_a^{(\alpha)})$
is a local rotation in the $xz$ plane of the Bloch sphere acting on qubit $a'$. $C$ maps $\ket{00}\rightarrow \ket{00}$, $\ket{01}\rightarrow \ket{11}$, $\ket{10}\rightarrow \ket{10}$ and $\ket{11}\rightarrow \ket{01}$ and the form of the local rotation is
\begin{equation}
R_{a'}(\theta_a^{(\alpha)})=\begin{pmatrix}
                          c_a^{(\alpha)} & -s_a^{(\alpha)} \\
                           s_a^{(\alpha)} & c_a^{(\alpha)} \\
                        \end{pmatrix}.
\end{equation}
Substituting this decomposition in to eq.~(\ref{eq:expectationvalN}) and expanding the terms gives
\begin{widetext}
\begin{equation}
\langle A^\alpha B^\beta \ldots Z^\omega \rangle =
\text{tr}\left[\sigma^{(3)}_a \sigma^{(3)}_b \ldots \sigma^{(3)}_z  C_{a'a}C_{b'b}\ldots C_{z'z}
\rho_{ab\ldots z}\chi_{a'b'\ldots z'}(\theta_a^{(\alpha)},\theta_b^{(\beta)},\ldots,\theta_z^{(\omega)})C_{a'a}C_{b'b}\ldots C_{z'z} \right].
\end{equation}
\end{widetext}
We have taken the local rotations $R(\theta)$ into the ancilla state so that $\chi_{a'b'\ldots z'}
(\theta_a^{(\alpha)},\theta_b^{(\beta)},\ldots,\theta_z^{(\omega)})= R_{a'}R_{b'}\ldots R_{z'} \chi_{a'b'\ldots z'}
(R_{a'}R_{b'}\ldots R_{z'})^\dag$. Using the cyclic property of the trace we can form $C_{a'a}
\sigma_a^{(3)} C_{a'a}=\sigma_a^{(3)} \sigma_{a'}^{(3)}$ for each of the parties. Substitution of this identity
allows one to break the trace up into two terms, one for the initial system state and one for the ancilla
state. The expectation value reads
\begin{equation}\label{eq:expectationvalgen}
\begin{split}
\langle A^{(\alpha)} B^{(\beta)} \ldots Z^{(\omega)} \rangle = \text{tr}\left[\sigma^{(3)}_a \sigma^{(3)}_b \ldots
\sigma^{(3)}_z \rho_{ab\ldots z}\right] \times \\ \text{tr}\left[\sigma^{(3)}_a \sigma^{(3)}_b \ldots
\sigma^{(3)}_z \chi_{a'b'\ldots z'}(\theta_a^{(\alpha)},\theta_b^{(\beta)},\ldots,\theta_z^{(\omega)})\right].
\end{split}
\end{equation}
The first term is just the expectation value $\langle \sigma_a^{(3)} \sigma_b^{(3)} \ldots \sigma_z^{(3)}
\rangle$ on the initial system state $\rho_{ab\ldots z}$ and is unaffected by the local angles
$\theta$. The second term is the expectation value of the
observables $\langle A^{(\alpha)} B^{(\beta)}
\ldots Z^{(\omega)} \rangle$ on the ancilla. The initial system state expectation value therefore just acts as
a multiplying constant on each local angle dependent ancilla expectation value. So iff
$\rho_{ab\ldots z}$ is an eigenstate of $\sigma_a^{(3)}
\sigma_b^{(3)} \ldots \sigma_z^{(3)}$ then $\rho_{ab\ldots z}(\theta_a^{(\alpha)}
\theta_b^{(\beta)}, \ldots, \theta_z^{(\omega)})$ violates
a Bell inequality by some amount iff the ancilla state $\chi_{a'b'\ldots z'}$ violates the same Bell inequality by the same amount. $\Box$

\emph{Comment:} In SKKVBs scheme, they chose the initial system state $\rho_{ab}=1/2(\matx{00}
+\matx{11})$ which is an eigenstate of $\sigma_a^{(3)}\sigma_b^{(3)}$ with eigenvalue $1$, a
special case of our more general conditions. Note that violation of a Bell inequality can still result even if the expectation value of the initial system state $\text{tr}[\sigma_a^{(3)} \sigma_b^{(3)}\rho_{ab}]
\neq \pm 1$. One just needs to ensure that the product of the ancilla and system state expectation values is greater
than the inequality bound.

As a simple example of a non-maximally entangled
ancilla state allowing violation of the CHSH inequality within this scheme we choose the classical product system state $\rho_{ab}=\matx{00}$
for simplicity, an eigenstate of $\sigma^{(3)}\sigma^{(3)}$ with eigenvalue $1$ and the Werner state
$\chi_{a'b'}=(1-p)\mathbf{I}/4+p\matx{\Psi^-}$ as the ancilla. It is well known that for $p>1/\sqrt{2}$
$\chi_{a'b'}$ violates the CHSH inequality \cite{ref:Werner89}.
One finds that, within this scheme, the separable reduced
state violates the CHSH inequality for the same conditions, namely $p>1/\sqrt{2}$. The system state is $\rho_{ab}(\theta_a^{(\alpha)},\theta_b^{(\beta)})= 1/2[(1-p\cos(\theta^{(\alpha)}_a-\theta^{(\beta)}_b))
\rho_{ab}+ (1+p\cos(\theta^{(\alpha)}_a-\theta^{(\beta)}_b))\bar{\rho}_{ab}]$ giving the expectation values $\langle A^{(\alpha)} B^{(\beta)} \rangle = -p\cos(\theta_a^{(\alpha)}-\theta_b^{(\beta)})$. The values of the local angles
giving maximal violation are $(\theta^{(1)}_a,\theta^{(2)}_a,\theta^{(1)}_b,\theta^{(2)}_b)=(0,\pi/2,\pi/4,-\pi/4)$.

\section{Multiparty Bell inequalities with entangled devices}\label{sec:examples}

We now give an example of violation of the Mermin
inequality \cite{ref:Mermin90a}, another particular case of WWZB, using our generalized scheme. We start with the initial system state $\rho_{abc}=\matx{000}$ and the Greenberger-Horne-Zeilinger
ancilla state $\chi_{a'b'c'}=1/\sqrt{2}(\ket{000}+i\ket{111})$. One wishes to calculate the expectation value of the operator
\begin{equation}\label{eq:merminoperator}
A=\sigma^{(1)}_a \sigma^{(1)}_b \sigma^{(2)}_c + \sigma^{(1)}_a \sigma^{(2)}_b \sigma^{(1)}_c + \sigma^{(2)}_a \sigma^{(1)}_b \sigma^{(1)}_c - \sigma^{(2)}_a \sigma^{(2)}_b \sigma^{(2)}_c.
\end{equation}
on the state $\rho_{abc}(\vec{\theta}_a^{(\alpha)},\vec{\theta}_b^{(\beta)},\vec{\theta}_c^{(\gamma)})$. $\sigma^{(1)}$ and $\sigma^{(2)}$ are the Pauli $x$ and $y$ operators
respectively. A state with a local hidden variable model has expectation value $\langle A \rangle \leq 2$. In SKKVBs scheme the
local rotations $R$ are in the plane $xz$, that is elements of $SO(2)$, a group with single angle $\theta$. $\sigma^{(2)}$ cannot be reached from $\sigma^{(3)}$
using such rotations so we generalize to the full set of local rotations on the sphere, $SU(2)$, parameterized with three angles given by the vector $\vec{\theta}$.
From the cyclic property of the trace we can move the rotations $R(\vec{\theta})$ so they act on the measurements in eq.~(\ref{eq:expectationvalgen}). This is picture (i) described earlier in the text. The rotations that take $\sigma^{(1,2)}=R(\vec{\theta})^\dag \sigma^{(3)} R(\vec{\theta})$, i.e. those appearing in eq.~(\ref{eq:merminoperator}), are not difficult to find. One can verify that the reduced state
$\rho_{abc}(\vec{\theta}_a^{(\alpha)},\vec{\theta}_b^{(\beta)},\vec{\theta}_c^{(\gamma)})$ is separable and
$\langle A \rangle = 4$ thus violating the inequality.

\section{Entanglement structure in the full state}\label{sec:clusterstate}

We showed that the expectation value of the final state
$\varrho_{ab\ldots z a' b' \ldots z'}$ is equivalent to the product of the expectation values of $\rho_{ab\ldots z}$
and $\chi_{a'b'\ldots z'}$ but this does not say anything about the structure of the final state. Here we give
the form of the final state in SKKVBs protocol and characterize its entanglement. Starting again with $\rho_{ab}=\matx{00}$ and $\chi_{a'b'}=\matx{\Phi^+}$ the final state reads
\begin{eqnarray}
&&\ket{\psi}_{ab\ldots z a' b' \ldots z'}=\\ &&\cos\left(\tfrac{\theta_a^\alpha-\theta_b^\beta}{2}\right)\tfrac{\ket{0000}+\ket{1111}}{\sqrt{2}} +
\sin\left(\tfrac{\theta_a^\alpha-\theta_b^\beta}{2}\right)\tfrac{\ket{1010}-\ket{0101}}{\sqrt{2}}\nonumber
\end{eqnarray}
which is equivalent under stochastic local operations and classical communication (and permutations of the qubits)
to the four qubit cluster state $1/2(\ket{0000}+\ket{0011}+\ket{1100}-\ket{1111})$, a class well known from the one-way model of quantum computation
\cite{ref:Raussendorf&Briegel01}. The entanglement structure of cluster states
is well known. An $N$ qubit cluster state remains entangled until $N/2$ of the qubits have been removed \cite{ref:Briegel&Raussendorf01}.
Thus the OLTs never leave $\rho_{ab}(\theta_a^\alpha,\theta_b^\beta)$ in an entangled state.

One can also verify that there are regions in phase space where the Wigner function of the overall cluster state is negative. By mapping the states $\ket{0}\rightarrow \ket{\alpha}$ and $\ket{1}\rightarrow \ket{-\alpha}$ one can make a straight forward although long calculation of the four mode Wigner function \cite{ref:Wigner32,ref:Hillery84}. When the entanglement in the devices is taken into account one finds the less surprising result that negativity of the Wigner function and entanglement is found in SKKVBs scheme.

\section{Conclusion}

In this \paper\ we have given a multiparty generalization of SKKVBs scheme and showed a necessary and sufficient condition on the initial system and ancilla states for a violation of a Bell inequality with dichotomic observables within these scheme. We have given some clarifying comments regarding SKKVBs scheme, namely that a Bell inequality may be violated by a classical system because one has violated the assumption of statistical independence through the dependence of the classical system on the free parameters of the measurement settings. The dependence on these free parameters was mediated by the entanglement in the ancillary system. We also showed that by switching your point of view that you can regard the scheme as a usual Bell test on the entangled ancilla. Once viewed in this way, our main result may seem less surprising.
 
However, our key message is that the violation of a Bell inequality does not necessarily imply that there was nonlocality in the system the Bell test was performed on. In fact the system being tested could have been in a classical state. In order for a violation to be observed there must have been nonlocality somewhere if one allows the measurement settings to be external free parameters. This nonlocality may have resided in an ancilla; possible candidates include the detectors or the environment. Concluding that the system being tested must have been entangled may be erroneous.

Our scheme also provides a method to determine whether any background, reference system may be nonlocal prior to performing a traditional Bell test.  One should take two separable systems, each located in one of the separate regions where the traditional Bell test would take place. OLTs (the unitary of eq.~(\ref{eq:unitary})) should then be applied between the system and background reference, which could, for instance, be the reference lasers in \cite{ref:Rowe01}. If a violation of a Bell inequality is found one can conclude that the background reference is nonlocal and may give rise to false violations in any traditional Bell test that takes place with the same set up.

These findings may be relevant experimentally. For example a recent nuclear magnetic resonance experiment testing a variant of the Bell-Kochen-Specker theorem \cite{ref:Kochen&Specker67,ref:Mermin93} used a very similar scheme to the one presented here \cite{ref:Moussa09}.

\emph{Acknowledgments} --- MSW and WS thank the National Research Foundation and the Ministry of Education (Singapore) for funding and LH acknowledges the EPSRC (UK) for support. We would also like to thank Jonathan Jones, Mauro Paternostro, Valerio Scarani, Vlatko Vedral and an anonymous referee for useful comments.

\bibliography{../../PhD_write_up/references/masterbib}

\end{document}